# Thin film of Al-Ga-Pd-Mn quasicrystalline alloy

T.P.Yadav

Hydrogen Energy Centre, Department of Physics (Centre of Advanced Studies),
Banaras Hindu University, Varanasi-220005, INDIA

## ABSTRACT

Thin film quasicrystal coatings have unique properties such as very high electrical and thermal resistivities and very low surface energy. A nano quasicrystalline thin film of icosahedral Al-Ga-Pd-Mn alloy, has produced by flash evaporation followed by annealing. The icosahedral phase of $Al_{65}Ga_5Pd_{17}Mn_{13}$ alloy has been used as a precursor material. The X-ray diffraction and transmission electron microscopy confirmed the formation of icosahedral phase in the thin film. The Energy dispersive X-ray analysis investigations suggest the presence of Ga (~ 5 at %) in the alloy. Icosahedral Al–Ga–Pd–Mn thin film provides a new opportunity to investigate the various characteristics including surface characteristics. The formation of icosahedral thin film in Al-Ga-Pd-Mn quaternary alloy by present technique has been studied for the first time. These films can be utilized as conversion coatings for Al substrates or incorporated into a full coating system containing an organic primer and a topcoat. Attempts will be made to discuss the micromechanisms for the formation of quasicrystalline thin film in Al-Ga-Pd-Mn alloys

Corresponding author: E-mail: yadavtp@gmail.com
Phone: 0542 2307307   fax: 0542 2368468

# 1. Introduction

Quasicrystals are orientation-ordered structures with classically forbidden rotation symmetries (e.g. fivefold and tenfold rotation axes) which are incompatible with periodic translational ordering. Quasicrystalline materials exhibit properties that are very different from conventional metallic materials; for reviews of their properties [1-2]. Industrial applications have employed quasicrystalline materials as a thin coating on conventional crystalline materials [3]. Therefore, it is important to establish the simple methods to grown the quasicrystalline thin film. Research in this area had been limited to know the formation of quasicrystalline thin film.

In Al- Pd- Mn alloy system, three types of quasicrystals, namely a stable decagonal phase with 1.2 nm periodicity, a stable icosahedral phase, and a metastable decagonal phase with 1.6 nm periodicity, successively appear with increasing Pd content in composition $Al_3(Mn_{1-x}Pd_x)$ (Tsai *et al*. 1989, Beeli *et al*. 1991, Hiraga *et al*.1991) [4-6]. Experimental results show that i-phase in Al-Pd-Mn alloy belongs to the face-centered icosahedral (FCI), similar to that in Al-Cu-Fe alloy and it remains stable up to the melting point. A thermodynamically stable approximant phases have been found in the compositional vicinity of quasicrystals in a number of quasicrystal-forming alloy systems. In the Al-Pd-Mn alloy system, besides icosahedral and decagonal phases, some orthorhombic (ξ') phases have been reported, which were shown to be quasicrystal approximants [7]. The stability range of the ξ phase, which has a structure closely related to that of the binary $Al_3Pd$ phase (Matsuo and Hiraga 1994) [8], is located in the low-Mn region. Considering the local structure of the material, ξ' (Al-Pd-Mn) is recognized to be an approximant of the well-known icosahedral phase $Al_{70.5}Pd_{21}Mn_{8.5}$. In particular, the icosahedral phase in Al-Pd-Mn has Mackay-type clusters as elementary structural building blocks (Boudard *et al*. 1996) [9]. The space group of the ξ' is P*nma* and the unit cell containing 316 atoms has lattice parameters $a = 23.54$ Å, $b = 16.56$ Å and $c = 12.34$ Å.

The formation of quasicrystalline phase in Al-Ga-Pd-Mn system is difficult due to the low melting temperature of Ga (29.76°C). A quaternary icosahedral *i*-Al-Ga-Pd-Mn quasicrystal (QC) with composition $Al_{67}Ga_4Pd_{21}Mn_8$ was discovered in 1999 by Fisher and co-workers,[10] following attempts to grow ternary *i*-Al-Pd-Mn from a fourth element flux in this case Ga.

Selected area diffraction patterns (SADPs) suggested that this new QC is isostructural with *i*-Al-Pd-Mn, where 4 at % Ga gets incorporated into the lattice. The quasilattice dimension of *i*-Al-Ga-Pd-Mn was found approximately 1.5% larger than that of the ternary compound. Apart from the original work, [11-12] the investigations of the structure and physical properties of the *i*-Al-Ga-Pd Mn QC are still scarce; therefore some basic issues remain open. It is not clear whether the Ga atoms occupy specific sites in the lattice or the Ga-Al substitution can be at random, i.e., whether this is a true quaternary or a pseudoquaternary QC. In addition, the reported electrical resistivity measurements suggest that as little as 4% Ga in the *i*-Al- Pd-Mn structure increases the resistivity by about 50%. The formation of QC phase in Al-Ga-Pd-Mn alloys system by simple melting and annealing processes has not been reported. The quasicrystalline thin film in Al-Ga-Pd-Mn system will be very important due to low surface energy of Ga (600mJ/m2) with respect to Al (1200 mJ/m2)

In the present paper, the formation of icosahedral quasicrystalline thin film in $Al_{65}Ga_5Pd_{17}Mn_{13}$ alloy by flash evaporation followed by annealing process will be reported with an aim to understand the microstructural evolution and phase stability during annealing.

## 2. Experimental details

The alloy with nominal composition of $Al_{65}Ga_5Pd_{17}Mn_{13}$ were prepared in an argon atmosphere by melting high purity Al (99.96%), Ga (99.99%), Pd (99.96%) and Mn (99.99%) in a silica crucible using RF induction furnace The individual elements were taken in correct stoichiometric proportions and pressed into a cylindrical pellet of 1.5 cm diameter, 1 cm thickness by applying a pressure of ~ $2.76 \times 10^4$ N/m$^2$. The pellet (5g by weight) was then placed in a silica tube surrounded by an outer Pyrex glass jacket. Under continuous flow of argon gas into the silica tube, the pellet was melted using radio frequency induction furnace (18 kW). During melting process, water is circulated in the outer jacket around the silica tube to reduce the contamination of the alloy with silica tube and also for good cooling. The melting atmosphere was purified by previously melting Ti buttons. The as-cast ingot was melted several times to ensure homogeneity. The homogenized alloy was then subjected to annealing at 700 & 800 °C. Before annealing the as cast ingots were sealed in a silica tube, which was flushed two to three times with high purity argon gas and then evacuated to a pressure of ~ $1.32 \times 10^{-6}$ atm, It was then placed in a furnace and annealed at 600°C ~ 900°C (±10°C) for time spans ranging from 10 to 80 h. It was found that annealing at 800 °C for 60 h produced the optimum material. Therefore, this time period was maintained for all further annealing runs. The optimum material was used for flash evaporation. The flash evaporation setup was developed by us in the department of physics, Banaras Hindu University see fig. 1. The as cast, heat-treated alloy and thin film was characterized by powder X-ray diffraction (XRD) using a Philips 1710 X-ray diffractometer with CuKα radiation (λ=1.54026Å). The surface microstructure of as cast and annealed alloy was characterized through scanning electron microscopy (SEM) (Philips: XL 20). The phase transformation temperatures were made with Linesis L72 differential thermal analyzer (DTA) in vacuum at heating rate of 10°K/S. The transmission electron microscopy (TEM) using Philips EM CM-12 has been used for microstructural and structural characterizations with an operating voltage at 100 kV. An energy-dispersive X-ray link with HRTEM Tecnai 20 $G^2$ system was employed for the compositional analysis.

## 3. Results and Discussion

Three Cu crucibles was used as materials cup with rotator and Tungsten (W) crucible was used for heating boat see fig 1.

In order to explore the surface morphology of as cast, annealed and thin film (as grown and annealed) of $Al_{65}Ga_5Pd_{17}Mn_{13}$, scanning electron microscopy (SEM) has been carried out. Fig2(a) shows SEM image of the as cast $Al_{65}Ga_5Pd_{17}Mn_{13}$ alloy. It can be seen that there are two types of contrast with spherical microstructure (marked as A, & B). A consists of long and straight rod like growth with diameter ranging from 30 to 40 of micrometers. However, the spherical type of growth on the surface of the rod with lengths from 6 to 10 of micrometers can be seen in fig 2(a). The morphology of fractured surface was obviously different from spherical microstructure. After 60 h of annealing at 800°C of as cast alloy a very smooth dodecahedran like microstructure has been observed (shown in fig. 2(b)). The dodecahedral microstructure of various quasicrystlline materials have been described extensively in the literature [13]. Therefore, it can be concluded that these small grains are icosahedral quasicrystalline phase. In addition to dodecahedral grain there are some grain having polygonal/petal morphology as well. Fig.2 (c) shows the SEM microstructure of as grown thin film, a fine cluster of particle can be seem. After annealing of the as grown thin film, the fine cluster was looking smooth and a bigger cluster was visible in fig. 2(d)

Figure 3(a) shows the XRD pattern of as grown $Al_{65}Ga_5Pd_{17}Mn_{13}$ thin film indicating the formation of icosahedral phase. An amorphous like hump around 2θ=25° has been noticed, this is due to the amorphous glass slide. The gradual formation of I phase after annealing for 60 h and 120 has been shown in fig 3(b-c). There are significant changes in XRD pattern after annealing of as grown thin film up to 120 h, at both the temperatures 300 °C which can be inferred from variation in peak intensity and broadening. After 120 h of annealing at 300°C, IQC phase has been observed (fig.3c) as dominant one. It was found that amorphous hump has disappeared after annealing; it is due to high intensity of the icosahedral phase. The XRD pattern of as cast and annealed alloy are not given for the sake of brevity.

The compositional analysis of as grown $Al_{65}Ga_5Pd_{17}Mn_{13}$ thin film was carried out by energy depressive X-ray analysis (EDX) attached to the transmission electron microscope Tecnai $G^{20}$ (shown in Fig.4).The EDX analysis shows the presence of all the elements in as cast alloy (Al= 63.9 at%, Ga= 4.2 at%, Pd= 20 at% and Mn= 11.9 at%) and it is very close to stoichiometric proportions. It also shows the very small presence of oxygen and Si contamination approximated to be around 0.15 and 0.6 at % in the sample. However, after annealing experiment the oxygen contamination increases up to 0.2 at % indicating the small oxygen pick up during annealing. It should be noted that the small oxygen pick up in this case is negligible in respect of quasicrystalline phase formation

To obtain further information on the phases mentioned above, we carried out transmission electron microscopic observations. Fig. 5(a-c) shows the typical selected area diffraction pattern (SADP) of annealed version (120 hours at 300 °C) of as-grown thin film which was taken from gray region of the microstructure (Fig.5d). Figure 5 (a-c) shows typical selected area electron diffraction patterns of the icosahedral quasicrystalline phase, taken along (a) two fold, (b) three fold axis, (c) mirror axis. In the corresponding microstructure the nano grain are clearly visible in Fig.5 (d). The results presented here suggest the existence of a field of stability for quaternary icosahedral quasicrystals thin film in Al-Ga-Mn-Pd at temperature of nearly 300°C and for compositions around $Al_{65}Ga_5Pd_{17}Mn_{13}$. The quasicrystal is formed when as grown thin film is annealed at 300°C for 120 h. In annealed specimens only icosahedral quasicrystal has been observed. This suggests that the icosahedral quasicrystal is stable in $Al_{65}Ga_5Pd_{17}Mn_{13}$ composition compared to the others crystalline phase.

## 4. Conclusion

In the present study we have obtained nearly pure icosahedral quasicrystalline thin film in Al-Ga-Pd-Mn by annealing the as grown thin film at 300 °C for 120 hours.

**Acknowledgement**

The authors would like to thank, Prof. O.N.Srivastava, Prof. R.S.Tiwari, Prof. N.K.Mukhopadhyay Prof. G.V.S.Sastry and Prof. R.K.mandal for many stimulating discussions. The financial support from CEFIPRA is gratefully acknowledged.

# Figure captions

**Fig.1** Experimental setup of flash evaporation technique developed at Physics Department, Banaras Hindu University.

**Fig.2** Scanning electron micrograph of (a) as-cast $Al_{65}Ga_5Pd_{17}Mn_{13}$ alloy, showing rod and spheroid like features which are marked A and B respectively (b) The microstructure obtained after annealing at 800 °C for 60 h, (c) as grown thin film (d) annealed version of thin film

**Fig.3** X-ray diffraction pattern obtained from (a) as grown thin film (b) annealed at 300° C for 60, (c) annealed at 300 °C for 120h.

**Fig.4** Energy–dispersive spectrum of the as grown $Al_{65}Ga_5Pd_{17}Mn_{13}$ thin film.

**Fig.5** The SADPs of annealed version of $Al_{65}Ga_5Pd_{17}Mn_{13}$ thin film, (a) two -fold (b) three fold (c) mirror plane and (d) corresponding bright field microstructure respectively.

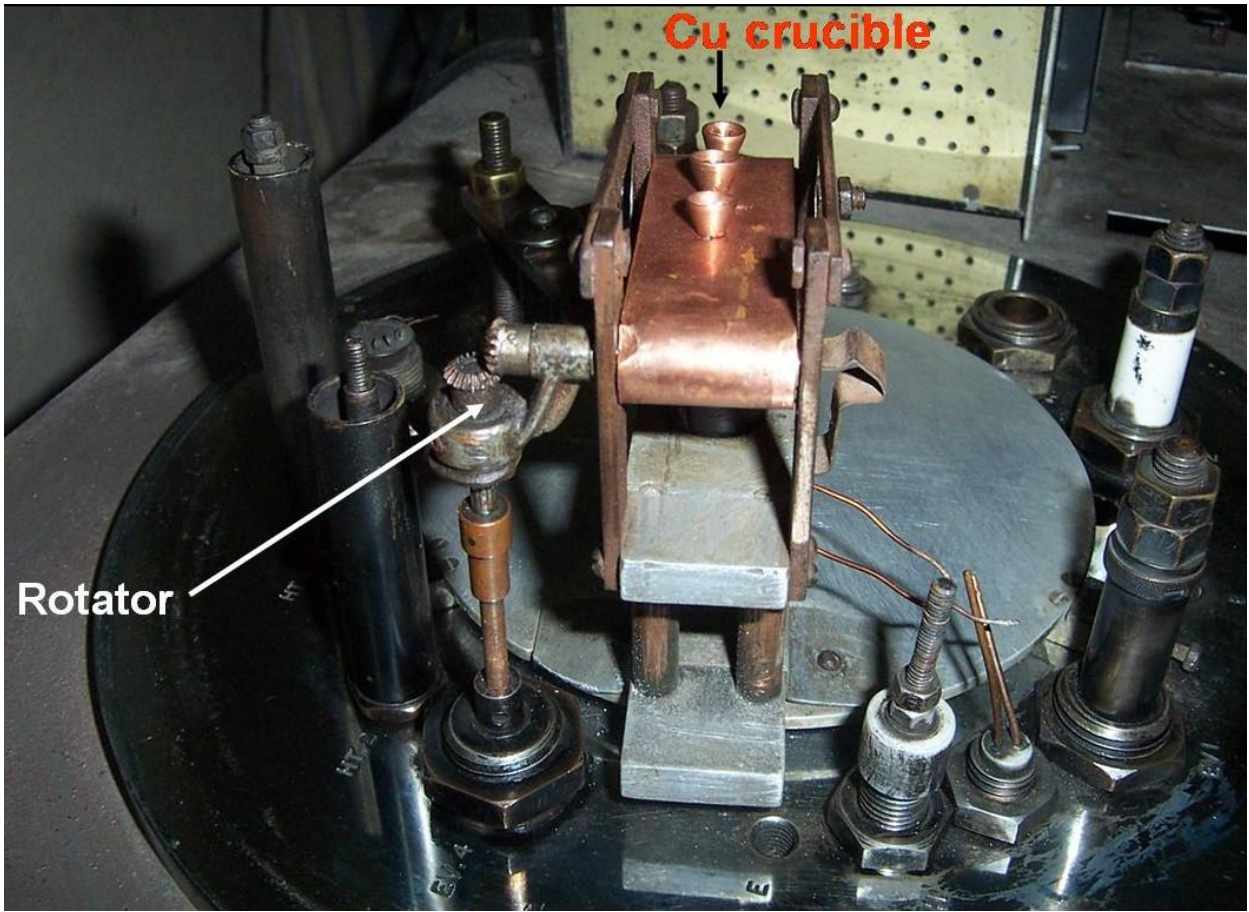

Fig.1

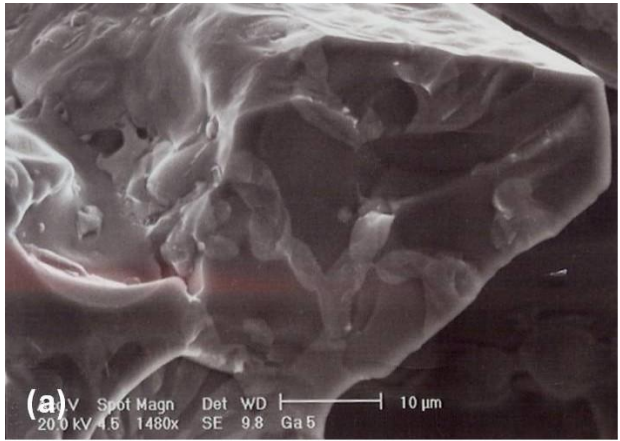 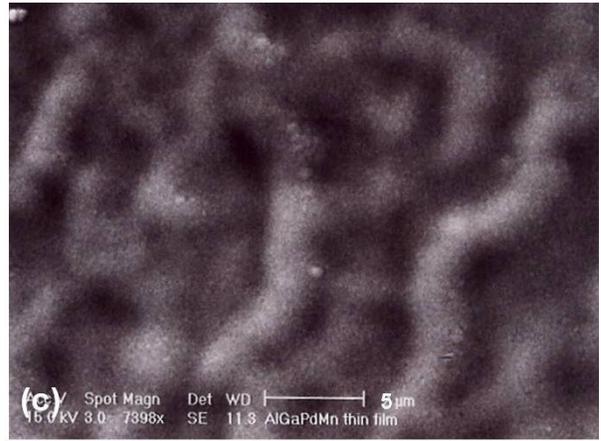
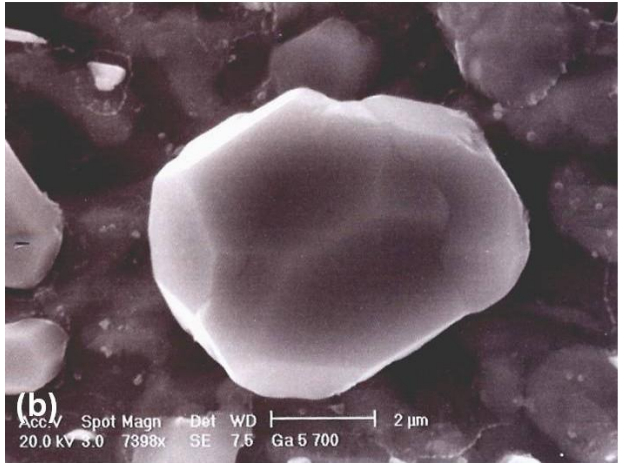 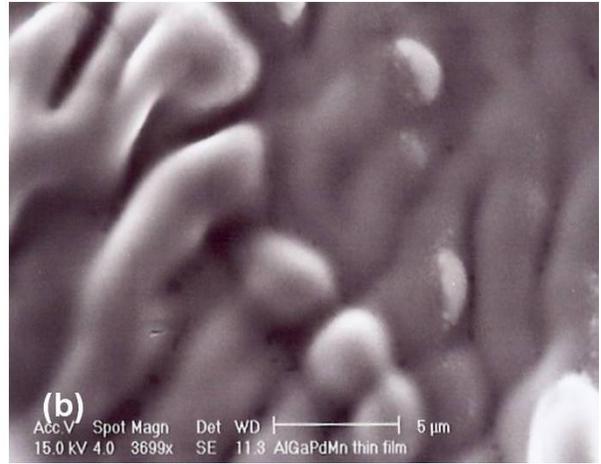

Fig.2

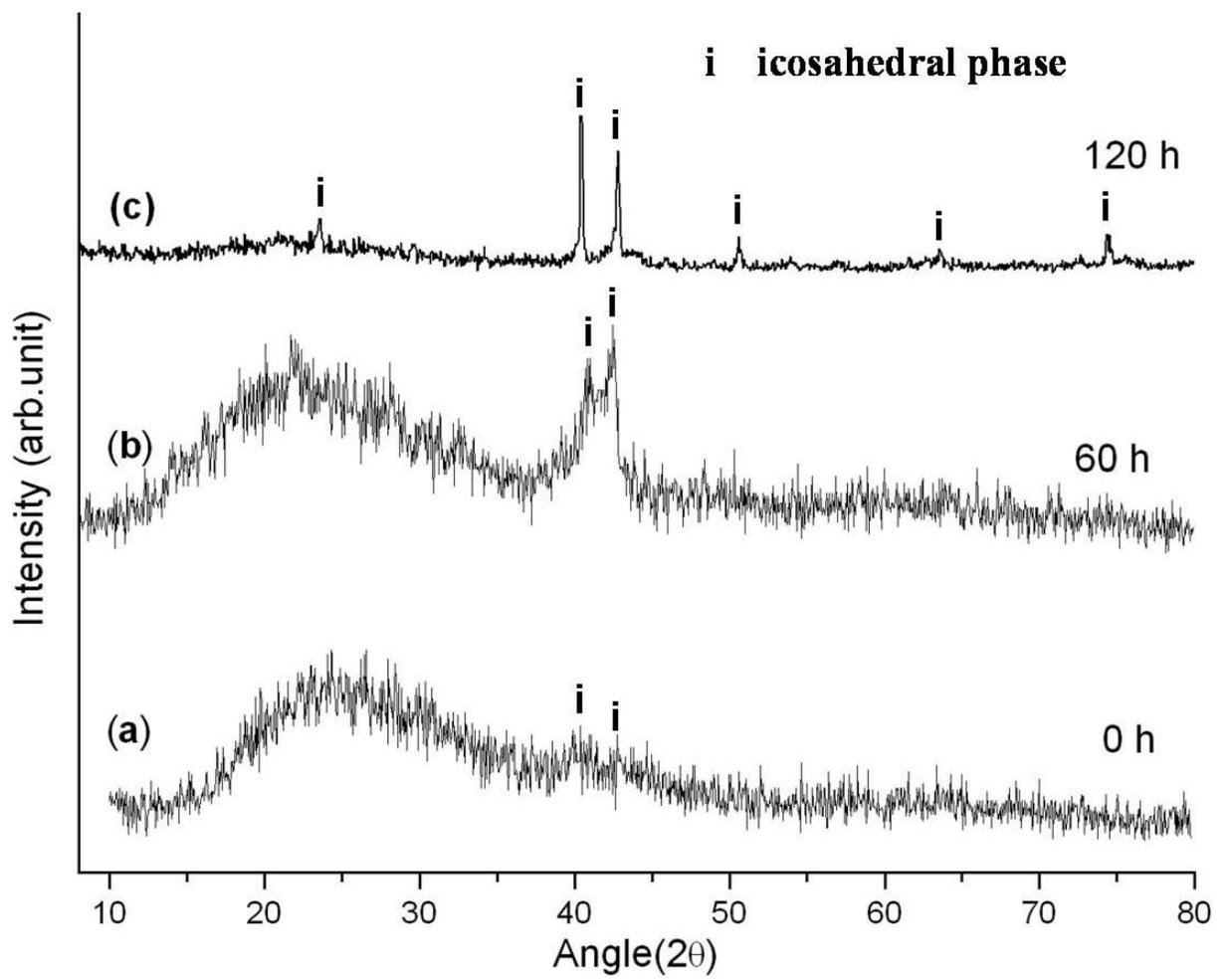

Fig.3

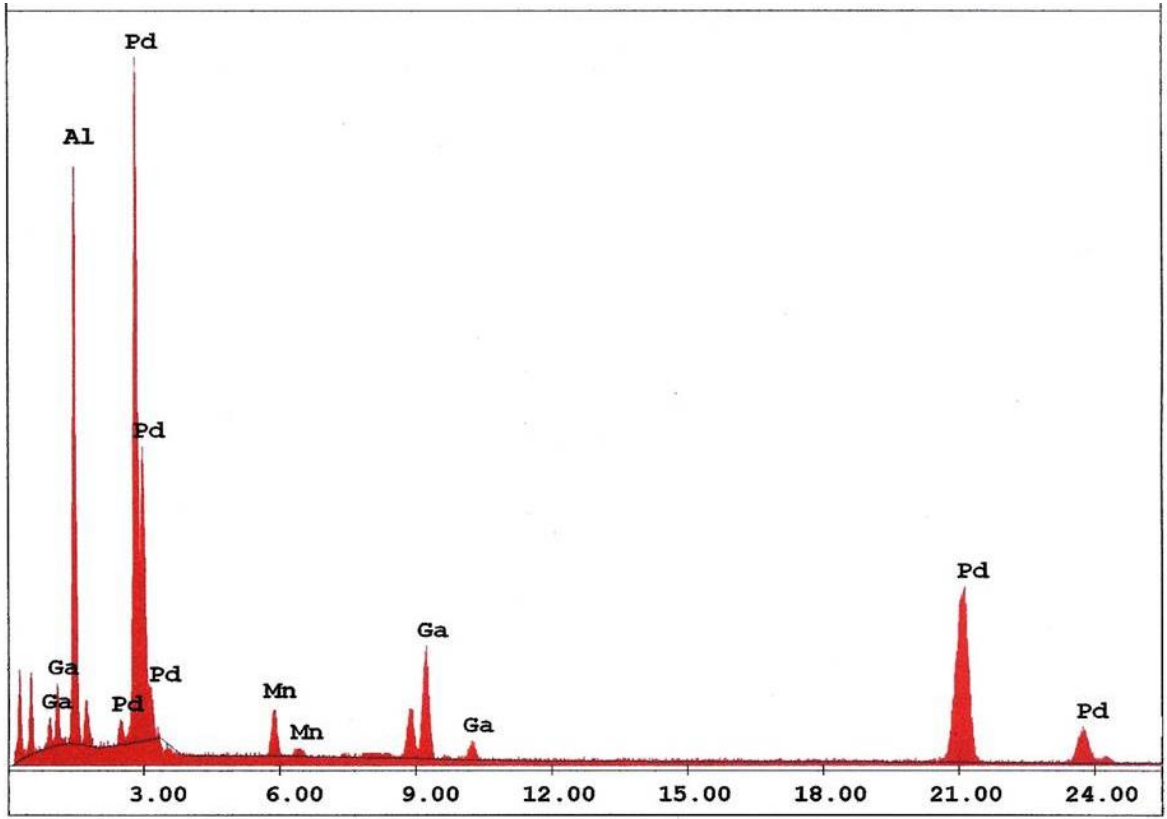

Fig.4

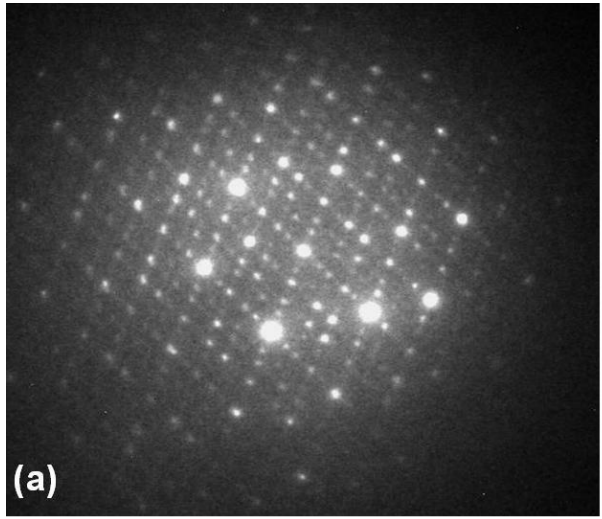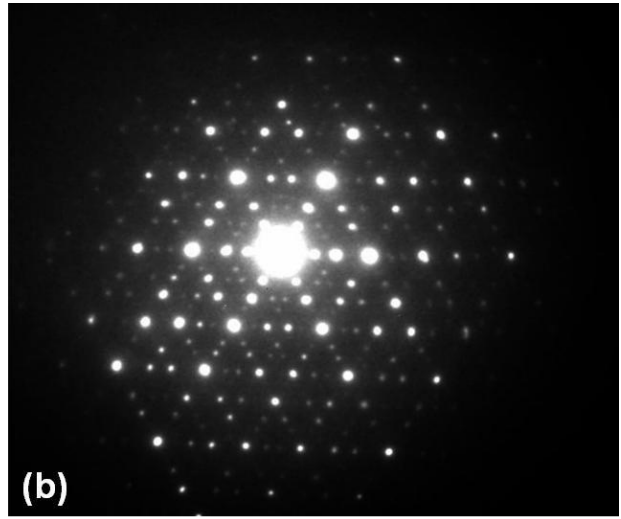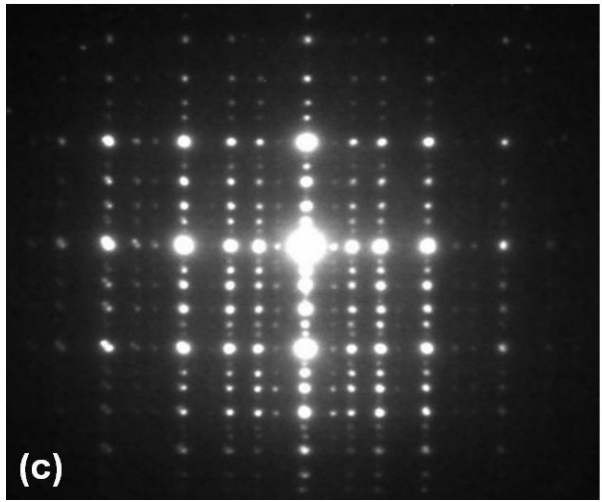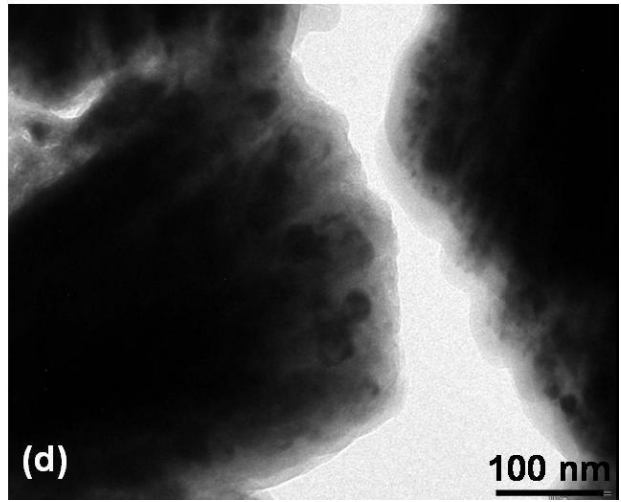

Fig.5